\documentclass[11pt,a4paper]{article}
\usepackage[hyperref]{acl2020}
\usepackage{times}
\usepackage{latexsym}
\usepackage{enumitem}
\usepackage{graphicx}
\usepackage{multirow}
\usepackage{subcaption}
\usepackage{booktabs}
\usepackage{mathtools}

\usepackage{microtype}

\aclfinalcopy

\newcommand{\var}[1]{\mbox{\emph{#1}}}

\newcommand{\Patk}[1]{\mbox{\Pat@$#1$}}
\newcommand{\RBPatp}[1]{\mbox{\RBP@$#1$}}

\newcommand{\NDCGatk}[1]{\mbox{\NDCG@$#1$}}
\newcommand{\ERRatk}[1]{\mbox{\ERR@$#1$}}

\newcommand{\svar}[1]{\mbox{\scriptsize\emph{#1}}}

\newcommand{\myurl}[1]{{\url{#1}}}
\newcommand{\mycaption}[1]{\caption{\normalfont{#1}}}

\newcommand{\myparagraph}[1]{\vspace{0.6\baselineskip}\noindent{\textbf{#1}}.~}

\newcommand{\mycomment}[1]{}

\newlength{\onedigit}
\newcounter{todocount}
 \definecolor{easy}{HTML}{1B9E77}
\definecolor{medium}{HTML}{D95F02}
\definecolor{hard}{HTML}{7570B3}

\title{RMITB at TREC COVID 2020}

\author{Rodger Benham \\
  RMIT University \\
  Melbourne, Australia \\\And
  Alistair Moffat \\
  The University of \\
  Melbourne, Australia \\\And
  J. Shane Culpepper \\
  RMIT University \\
  Melbourne, Australia \\
}

\date{}

\begin{document}
\maketitle
\begin{abstract}
    Search engine users rarely express an information need using the same
query, and small differences in queries can lead to very different
result sets.
These {\em user query variations} have been exploited in past TREC
CORE tracks to contribute diverse, highly-effective runs in offline
evaluation campaigns with the goal of producing reusable test
collections.
In this paper, we document the query fusion runs submitted to the
first and second round of TREC COVID, using ten queries per topic
created by the first author.
In our analysis, we focus primarily on the effects of having our
second priority run omitted from the judgment pool.
This run is of particular interest, as it surfaced a number of
relevant documents that were not judged until later rounds of the
task.
If the additional judgments were included in the first round, the
performance of this run increased by $35$ rank positions when using RBP $\phi
= 0.5$, highlighting the importance of judgment depth and coverage
in assessment tasks.
 \end{abstract}

\section{Introduction}

Harnessing the variability of user queries for a topic to improve
search effectiveness has been validated in many studies.
{\citet{bmst16sigir}} created a test collection for ClueWeb12-B using
$100$ topics and $10{,}835$ collected query variations; and the recent
CC-News English newswire corpus has also supplied query variations
{\citep{mackenziecc}}.
Inspired by the double fusion experiments of {\citet{bmst17-sigir}},
experts solicited query variations for the Robust 2004
topics~{\cite{voorhees2004overview}} as well as a few new topics,
to be used for the RMIT~{\cite{benhamrmit}} runs
submitted to the TREC CORE 2017 track~{\cite{allan2017trec}}.
RMIT participated again the following year~{\citep{benhamrmit2}},
producing the second-best run by AP, using shallow judgments
to identify and fuse the most effective query variations.
In follow-up work, {\citet{bmmc19-tois}} show that query fusion can
be applied to support boosting effectiveness at query time using
CombSUM.

In this work, we apply query fusion to efficiently and
effectively retrieve answers to questions from the scientific
literature collected during the COVID-19 pandemic.
We dissect the decisions made between submission rounds, using
additional judgments gathered for topics appearing in previous
rounds.
In late 2019, Severe Acute Respiratory Syndrome Coronavirus 2
(SARS-CoV-2) emerged as a health and economic disaster worldwide.
The severity of the pandemic has led to a spike in scientific
publications about the COVID-19 disease caused by the virus,
prompting {\citet{wang2020cord}} to create the CORD-19 dataset in
order to encourage IR-related explorations into COVID-related
scientific outcomes.

TREC COVID is the first TREC track to use the {\em residual
collection scoring} pooling methodology described by
{\citet{salton1990improving}} (see also {\citet{mwz07sigir}}).
{\citet{voorhees2020effect}} found that the effectiveness of our
second first round run {\texttt{RMITBFuseM2}} increased $33$
positions in the overall system ranking when using P@5 and after
additional judgments were gathered for those original topics during
later rounds.
Here we continue this line of inquiry using a post-hoc
analysis and the full judgment set from all six rounds of the
challenge.
 \section{First Round}

The COVID-19 pandemic differs from previously explored IR contexts by
virtue of the rapidly changing new information published daily
over an extended period of time.
When combined with the wide variety of questions being asked about
the disease, creating a practical IR evaluation exercise with the
potential to produce valuable insights for future pandemics is a
challenge.
In the first round of the task RMIT submitted two double fusion runs.
The first run {\texttt{RMITBM1}} re-weighted relevance scores of
documents by freshness, and {\texttt{RMITBFuseM2}} served as a
control to determine the efficacy of that time-biased approach.
Since {\texttt{RMITBFuseM2}} contained a proper subset of the
techniques applied to {\texttt{RMITBM1}}, we first describe how that
run was built.

\myparagraph{Processing The Corpus}
The first round uses the CORD-19 dataset as at April 10, 2020, with
the commercial, non-commercial, custom license, and bioRxiv subsets.
The corpus includes a metadata CSV file with various attributes about
publications, with fields including but not limited to: the title,
authors, an abstract, and the filename of the associated PMC and/or
PDF JSON parse of each publication.

Participants were instructed to prefer the PMC parse over the PDF
parse.
Both of these parses were supplied in JSON, where the document text
is dispersed in JSON objects with metadata about the context of each
sentence in relation to its place in the document.
That level of detail is superfluous for the retrieval models we
employed, and we extracted the document text from these
objects and transformed each parsed document into plain-text.

The stipulated list of document identifiers for the first round
excluded previously judged documents in an initial pool of three
Anserini baselines judged to depth-$40$~\citep{voorhees2020trec}.
Many of the documents to be assessed did not have an associated PMC
or PDF parse, with an abstract only.
Since these records had no document to process we did not index them.
(We subsequently found this decision to be detrimental, as
abstract-only document records were judged in the first round).

We used Terrier to index the corpus, as {\citet{kc18-sigir}} show that
it can produce double fusion runs with higher effectiveness than the
best submitted run to Robust04.
{\citet{lin2020reproducibility}} recently showed that Terrier
configurations are the most reproducible out of a series of tests
applied to popular IR retrieval tools.

\myparagraph{Query Variations}
The TREC COVID task operated with short preparation windows.
Each team had roughly a week to submit runs after receiving judgments
for the prior round.
That short time-frame prevented the design of a crowdworker study to
solicit query variations.
Instead, the first author of the paper took on the task of creating
$10$ query variations for each of the initial set of $30$ topics.

For example the second topic was the query {\em coronavirus
response to weather changes}, with the associated narrative: 
\begin{quote}
Seeking range of information about the SARS-CoV-2 virus viability in
different weather/climate conditions as well as information related
to transmission of the virus in different climate conditions.
\end{quote}

After having read that narrative and interpreting the question in the
released topic set, these ten queries were created: 

{
\em
\begin{enumerate}[itemsep=1.5pt,parsep=0pt]
    \item coronavirus climate change
    \item coronavirus weather
    \item coronavirus humidity
    \item coronavirus cool dry climate
    \item coronavirus cool humid climate
    \item coronavirus viability cool temperatures
    \item coronavirus winter temperatures viable 
    \item coronavirus summer temperatures viable 
    \item coronavirus seasonal climate 
    \item coronavirus standard laboratory conditions
\end{enumerate}
}

\vspace{-1.5ex}
\myparagraph{Double Fusion}
{\citet{bmst17-sigir}} proposed double fusion as a technique to
submit query variations to many retrieval models and fuse the results
with a rank fusion algorithm.
The technique was used previously to create the second-most effective
TREC CORE 2018 adhoc run~\citep{benhamrmit2}.

Initially, the aim was to use $16$ different retrieval models with
and without query expansion, $32$ rankings in total.
However limitations in the fusion script, and the short time-lines
being worked to (meaning that software modifications were risky),
meant that only eight retrieval models were used in the first round.

Many Terrier retrieval models are derived from the divergence from
randomness (DFR) model~\citep{amati2002probabilistic}.
Of the available options, the following variants were used to form
rankings in the first round submission: 

\begin{itemize}[itemsep=1.5pt,parsep=0pt]
    \item BB2 (DFR) \citep{plachouras2004university}
    \item BM25 \citep{robertson1995okapi}
    \item DFR\_BM25 (DFR) \citep{amati03thesis}
    \item DLH (DFR) \citep{macdonald2005university}
    \item DLH13 (DFR) \citep{macdonald2005university}
    \item DPH (DFR) \citep{amati2008fub}
    \item DFRee (DFR) \citep{amati2011fub}
    \item Hiemstra\_LM \citep{hiemstra2001using}
\end{itemize}

Running the $300$ queries against the eight models with and without
expansion resulted in the generation of $4{,}800$ rankings to depth
$1{,}000$.
To fuse each of the $160$ rankings per-topic, CombSUM was then
employed {\citep{shaw1995combination}}, as described by
{\citet{bmmc19-tois}}.
The run {\texttt{RMITBFuseM2}} was the result of fusing across (for
each topic) systems and queries.

\begin{figure}[t!]
    \includegraphics{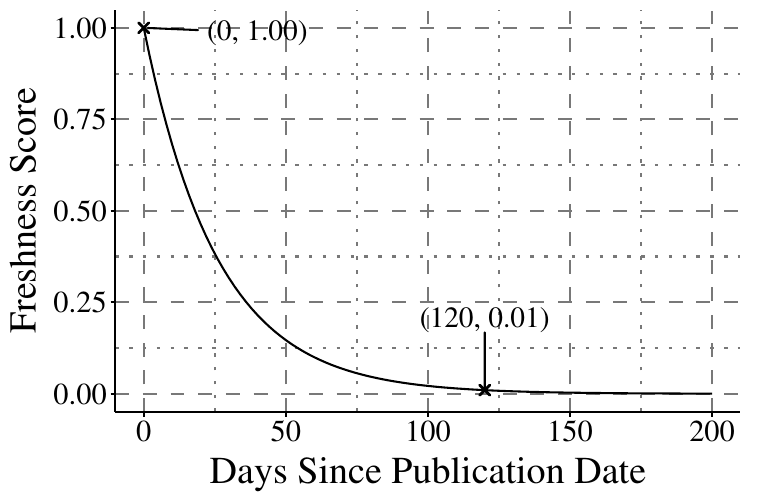}
    \mycaption{Experimental freshness scoring function employed in 
    \texttt{RMITBM1}. The function is an exponential decay
    through the two points marked with crosses. \label{fig-values}}
\end{figure}

\myparagraph{Freshness}
Knowledge of the COVID-19 pandemic is constantly evolving.
The second run, {\texttt{RMITBM1}}, was the result of combining a
freshness score with the orderings present in the
{\texttt{RMITBFuseM2}} run, based on the hypothesis that facts
disseminated four months ago are likely to be obsolete compared to an
article published on the same topic more recently.
With that, we parse the publication date out of the supplied metadata
file and calculate the number of days since the paper was published.
We then define an exponential decay on the variables $\var{days}$
since publication, fitting freshness scores through the two values:
$(0, 1)$ and $(120, 0.01)$.
{Figure~\ref{fig-values}} visualizes that fitted function, plotting
the derived formula:
\begin{equation}
    t(\var{days}) = 0.9623506264^{\svar{days}} \, .
\end{equation}

The publication dates in the supplied corpus metadata file required
cleaning.
Some articles were erroneously published in the future on the last
day of $2020$; those dates were adjusted to be the last day of
$2019$.
Other future dates corresponded to the date of a conference or when a
journal article was to be officially made available, but not when the
work was first disseminated.
In such instances, where the days since publication produced negative
integers, $\var{days}$ was set to $0$.
On rare occasions, the date format would change from {\texttt{Y-m-d}}
to {\texttt{Y},} and so, it would be parsed as the first day of that
year.
Empty date strings were taken to be the first day of $2020$.

To combine a freshness score with the relevance score of a document
assuming equal importance, the document scores in a run are adjusted
to be between $0$ and $1$ with minimax scaling to be cast in the same
units of freshness.

A simple unweighted linear combination is used, and so, the adjusted
document score is
\begin{equation}
    \var{score}(d, q) = {\mathcal{M}}(s_{q}, 0, 1) + t(\var{days}),
    \label{eqn-scoring}
\end{equation}
where $\mathcal{M}$ represents the minimax function, and $s_q$ refers
to the relevance score of the document in response to the query $q$.

After applying {Equation~\ref{eqn-scoring}} to
{\texttt{RMITBFuseM2}}, and sorting the document list of each topic
by the respective $\var{score}(d, q)$, we get {\texttt{RMITBM1}}.

\begin{figure*}
\centering
\begin{subfigure}{.49\textwidth}
  \raggedleft
    \includegraphics[width=1.0\columnwidth]{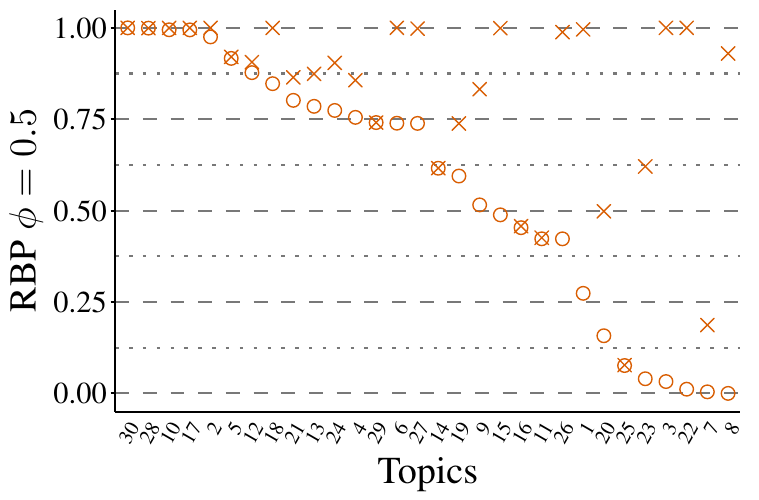}
    \caption{\texttt{RMITBFuseM2}}
  \label{fig-rbp05-pertopic-run2}
\end{subfigure}
\begin{subfigure}{.49\textwidth}
  \raggedright
    \includegraphics[width=1.0\columnwidth]{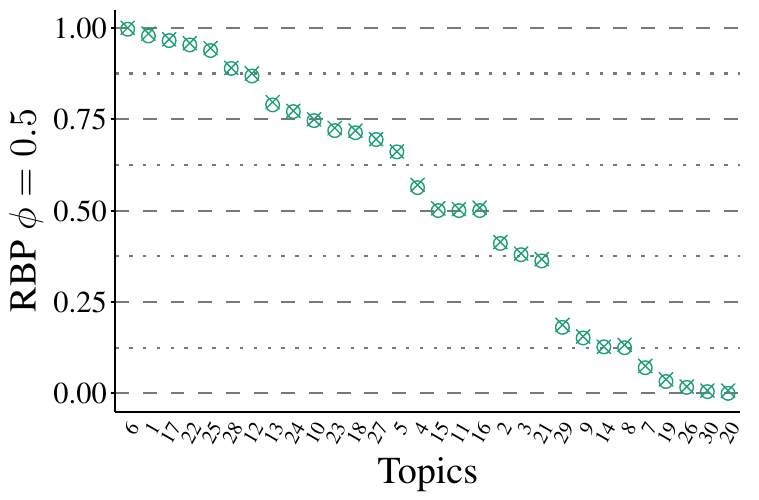}
    \caption{\texttt{RMITBM1}}
  \label{fig-rbp05-pertopic-run1}
\end{subfigure}\hfill
\caption{Per-topic RBP $\phi = 0.5$ evaluation of our submitted TREC
COVID round 1 runs.
Circles mark the RBP score, and the corresponding crosses mark the
``RBP plus residual'' score, indicating the maximum possible score
for that topic if all unjudged documents turned out to be relevant.
On the right, run {\texttt{RMITBM1}} contributed to the judgment pool
and has small residuals, whereas run {\texttt{RMITBFuseM2}} on the
left did not, and hence might potentially have identified additional
relevant documents and obtained an even higher mean RBP score.}
\label{fig-rbp05-pertopic}
\end{figure*}

\myparagraph{Analysis}
After the first round, the organizers shared the judgments and the
evaluation summaries of each submitted run.
Each run summary contained the measures: total relevant, total
relevant retrieved, average precision, mean BPref, and mean NDCG@10.
These measures are based on knowledge of how many relevant documents
there are in the collection, which when judging to depth $7$ in a
residual collection pooling context seems unlikely to provide
complete figures.
A graph showing deviation of median P@5 indicated that our
{\texttt{RMITBM1}} run was less effective than anticipated.

\begin{figure}[t!]
\centering
\includegraphics{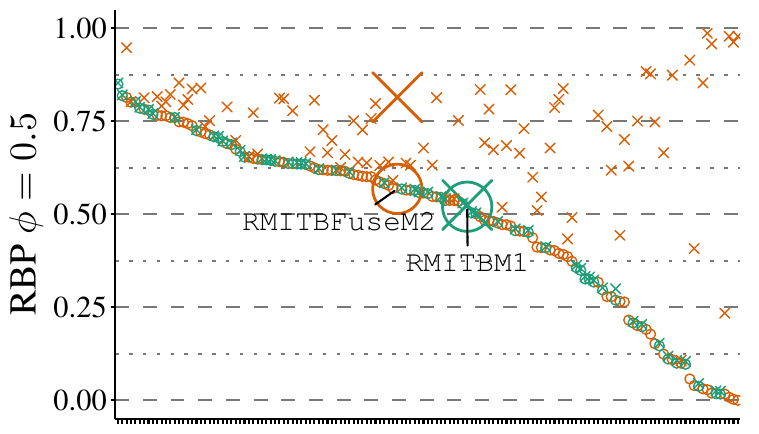}
\mycaption{RBP $\phi = 0.5$ evaluation of the $143$ runs submitted to
TREC COVID round 1, where circles mark the RBP score, and crosses
mark the corresponding ``RBP plus residual'' score.
Run {\texttt{RMITBM1}} is in the middle of the pack, and the unpooled
{\texttt{RMITBFuseM2}} run outperforms it, with the possibility of
having been within the group of top runs if the judgments provided
more coverage.
\label{fig-rbp05}}
\end{figure}

The organizers later reported rank-biased precision
(RBP)~{\cite{mz08acmtois}} in subsequent rounds with an expected
viewing depth of the top-$2$ documents ($\phi = 0.5$).
RBP does not rely on knowing the count of relevant documents for each
topic -- a statistic that has been argued to be untrustworthy for use
in pooled evaluation campaigns~\citep{zobel2009against}.
{Figure~\ref{fig-rbp05}} shows the monotonically decreasing RBP $\phi
= 0.5$ minimum score of each submitted run marked as a circle, with
the associated score uncertainty (the {\em residual}) marked with a
cross.
Points marked in green indicate that the system was pooled, and brown
points indicate otherwise, a convention used consistently throughout
the paper.
We were cautiously optimistic about the {\texttt{RMITBFuseM2}} run
after checking the RBP residuals, as it was more effective than
{\texttt{RMITBM1}}, even though it didn't contribute to the pool of
judgments.
Figure~\ref{fig-rbp05-pertopic} shows the monotonically decreasing
topic scores for each of our two submitted runs.
In Figure~\ref{fig-rbp05-pertopic-run2}, although the residual and
minimum score are close on some topics, for many there is a wide
diverge.
In particular, where the cross is near $1.0$, it means that none of
the document retrieved near the top of the run for that topic had
been judged to be {\emph{non}}-relevant either.

{Figure~\ref{fig-rbp05}} shows that the freshness reweighting
function in {Equation~\ref{eqn-scoring}} was not as useful as we had
anticipated.
Either there were many relevant documents older than four months, or
the freshness signal dominated the relevance signal in a way that
harmed effectiveness.
As only one run per-group had been judged $7$ documents deep, there
was no incentive to tune a coefficient in a linear combination of
relevance and freshness scores, as we risk overfitting our model to
reduce the diversity of our run, and submitting a control run might
not be judged.

With the uncertainty brought about by the shallow judgments, we now
move to discuss the decisions made in submitting our second round
run.
 \section{Second Round}
\label{sec-second-round}

In assessing risk aspects of the pipeline used in the poorly
performing round 1 submission of {\texttt{RMITBM1}}, we: 

\begin{itemize}[itemsep=1.5pt,parsep=0pt]
    \item remove freshness re-ranking;
    \item disable query expansion;
    \item add extra eight retrieval models; and
    \item include abstract-only documents.
\end{itemize}

Although query expansion is a powerful tool, it often requires
additional parameter tuning with relevance judgments to avoid query
drift, as these parameters vary across
corpora~\citep{billerbeck2004questioning}.
Noting the shallow judgments issue in the analysis provided in the
previous section, we abandoned query expansion, uncertain as to
whether it was reducing the effectiveness of the query fusion with
the default Terrier parameters.
We will instead explore this option in future work.

As previously discussed, our implementation was able to fuse up to
$16$ runs for each topic, and with expansion removed, we returned to
our original list of systems, adding in eight further alternatives: 
\begin{itemize}[itemsep=1.5pt,parsep=0pt]
    \item IFB2 (DFR) \citep{plachouras2004university}
    \item In\_expB2 (DFR) \citep{plachouras2004university}
    \item In\_expC2 (DFR) \citep{plachouras2004university}
    \item InL2 (DFR) \citep{plachouras2004university}
    \item LemurTF\_IDF \citep{zhai2001notes}
    \item LGD \citep{clinchant2009bridging}
    \item PL2 (DFR) \citep{plachouras2004university}
    \item TF\_IDF \citep{jones1972statistical}
\end{itemize}
Combining these eight models with the original eight, and not
employing query expansion, meant that there were again $16$ systems
being fused for each topic.

\begin{figure}[t!]
\includegraphics{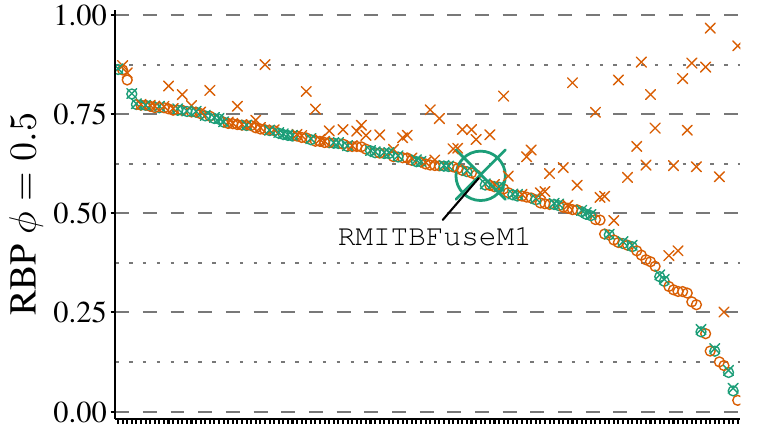}
\mycaption{RBP $\phi = 0.5$ evaluation of the $136$ runs submitted to
TREC COVID round 2, where circles mark the RBP score, and crosses
mark the RBP score plus residual.
Our \texttt{RMITBFuseM1} run generated only average performance.
\label{fig-rbp05-rnd2}}
\end{figure}

\myparagraph{Analysis}
{Figure~\ref{fig-rbp05-rnd2}} shows that the resulting round 2
{\texttt{RMITBFuseM1}} run was slightly below average in
effectiveness relative to other submitted runs.
{\citet{benhamrmit2}} found that a small pool of judgments used to
select the top-$5$ most effective query variations independently per
topic led to effectiveness improvements.
But the tight time constraints for TREC COVID, and a lack of medical
expertise, meant that undertaking manual judgments was not an option.

After inspecting the residuals shown
{Figure~\ref{fig-rbp05-pertopic}}, where {\texttt{RMITBFuseM2}} only
has one setting turned off compared to {\texttt{RMITBM1}}, it does
not appear that a failure analysis based on the run components will
be fruitful at this point.
 \section{Post-Hoc Analysis}

\begin{figure}[t!]
\centering
\includegraphics{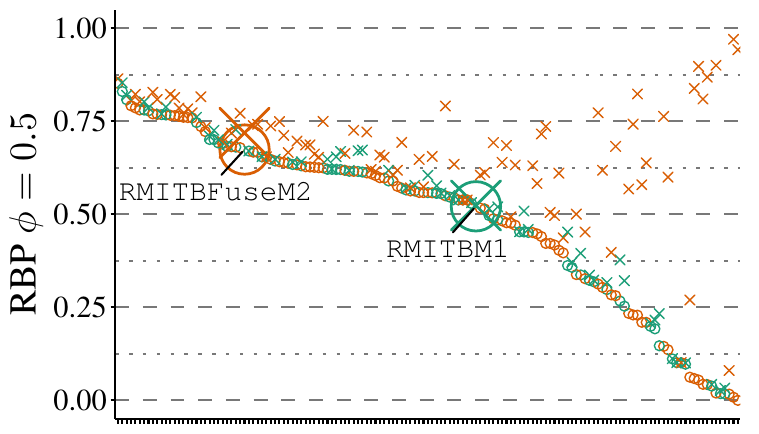}
\mycaption{RBP $\phi = 0.5$ evaluation of the $143$ runs submitted to
TREC COVID round 1 using the complete judgments provided at the end,
to be directly compared against Figure~\ref{fig-rbp05}.
Run {\texttt{RMITBFuseM2}} was originally ``average'', but moves
into the top-quartile.
\label{fig-rbp05-complete}}
\end{figure}

{\citet{voorhees2020effect}} provides a post-hoc analysis after the
complete judgments had been shared, showing that the relative system
orderings would have changed for {\texttt{RMITBFuseM2}}:, and noting:
\begin{quote}
  \em
  The largest change in the relative ranking of runs is the
  {\mbox{\normalfont\texttt{RMITBFuseM2}}} run which rises 33 ranks
  when using P@5 as the measure (21 ranks by NDCG@10, 7 ranks by MAP
  and none for BPref).
  \end{quote}
The RBP analysis shown in Figure~\ref{fig-rbp05} hinted that
{\texttt{RMITBFuseM2}} could have been in the top-third of first
round submissions, but all that could be concluded based on the
round~1 judgments was that it was at least average in effectiveness.
Figure~\ref{fig-rbp05-complete} plots the same first round systems,
but using the larger set of qrels, and shows that when measured by
RBP, {\texttt{RMITBFuseM2}} moves up $35$ places.
Enjoying the wisdom that comes with hindsight, if we had known that
outcome, we might not have disabled the query expansion features
mentioned in Section~\ref{sec-second-round} in our second round
submission.

\begin{figure}[t!]
  \centering
  \includegraphics{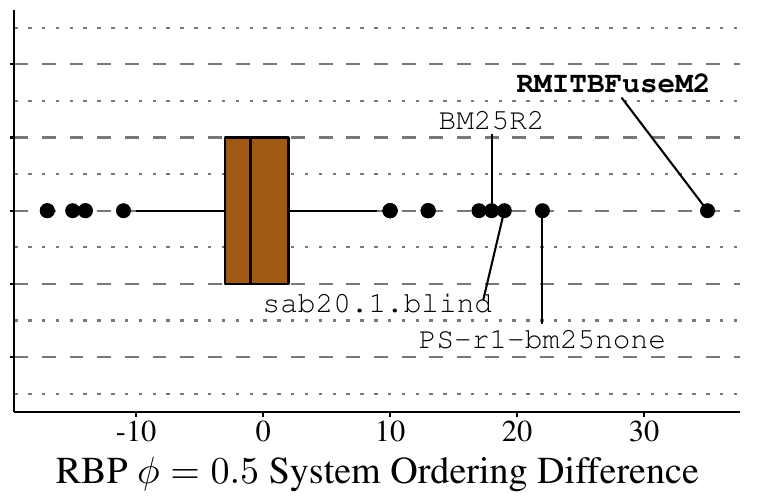}
  \mycaption{System rankings changes based on RBP $\phi = 0.5$
  evaluation using the round 1 judgments only, and then the complete
  set.
  Run {\texttt{RMITBFuseM2}} moves up the most places out of the
  submissions, where runs are labeled if they are extreme outliers
  ($3.0 \times \var{IQR}$).\label{fig-rbp05-rank-changes}}
  \end{figure}

\begin{table}[]
  \begin{tabular}{@{}lccc@{}}
\toprule
System & Rank & RBP & Residual \\ \midrule
\multicolumn{4}{c}{Round 1 Judgments}      \\ \midrule
\texttt{RMITBFuseM2}      &
    65 & 
        0.586 & 
            0.246 \\
\texttt{PS-r1-bm25none}      &
    60 &
        0.592 & 
            0.205 \\
\texttt{sab20.1.blind}      &                  
    74 & 
        0.547 & 
            0.267      \\
\texttt{BM25R2}      &                  
    91 & 
        0.468 & 
            0.366      \\ \midrule
\multicolumn{4}{c}{Complete Judgments}     \\ \midrule
\texttt{RMITBFuseM2}      &                  
    30 & 
        0.674 & 
            0.045      \\
\texttt{PS-r1-bm25none}      &                  
    38 & 
        0.641 & 
            0.108      \\
\texttt{sab20.1.blind}      &                  
    55 & 
        0.615 & 
            0.110      \\
\texttt{BM25R2}      &                  
    73 & 
        0.557 & 
            0.098      \\ \bottomrule
\end{tabular}

   \mycaption{Comparing the effectiveness of the extreme outlier runs
  for RBP $\phi =0.5$ rank changes shown in
  Figure~\ref{fig-rbp05-rank-changes}, where rank refers to the RBP
  system ranking measured against the first round
  runs.\label{tbl-extra-judgments-rbp} }
\end{table}

Figure~\ref{fig-rbp05-rank-changes} shows a boxplot of the relative
ranking changes of all systems on the RBP $\phi = 0.5$ measure from
the smaller first round judgment set to the full set.
Most systems have modest changes in rank, however, there are
outliers, with the extreme outliers labeled on the graph.
Of these outliers: \texttt{PS-r1-bm25none} also generated a query
manually from the topic descriptions; \texttt{sab20.1.blind} is a
pseudo relevance feedback run without abstracts; and \texttt{BM25R2}
is a BM25 run where the index contains the title, abstract, and
paragraph fields combined with Anserini's {\em CovidQuery Generator}
to generate queries.

Table~\ref{tbl-extra-judgments-rbp} documents these RBP-based
relative system orderings, along with the effectiveness scores and
residuals for the first round judgment set compared with the complete
judgment set.
Although the residuals are smaller on the complete judgment set, it
is possible that \texttt{PS-r1-bm25none} or \texttt{sab20.1.blind}
could outrank our \texttt{RMITBFuseM2} run with complete judgments,
and could, potentially, result in further jumps in system ordering of
$8$ and $25$ places respectively.

 \section{Conclusion}

We have documented our participation in the TREC COVID track.
While early evaluation outcomes in the per-round analysis indicated
that our runs were at best average, deeper judgments on the first
round run {\texttt{RMITBFuseM2}} lifted its system ranking by $35$
positions (RBP $\phi = 0.5$).
We found the residual pooling approach to be a refreshing take on
judgment solicitation with feedback, and welcome a similar approach
being applied to future evaluation campaigns.
Our recommendation would be to have fewer rounds and allow for more
judgments per round, where each feedback round is evaluated with
fixed pooling conditions.
Not only would this reduce the volatility in judgment coverage
observed in this year's track, it would also provide a higher quality
test collection for detailed failure analyses of system submissions
in all rounds.
We also encourage the use of residuals as a way of gauging the extent
to which measurements derived from pooled judgments can be considered
to be reliable.

\section*{Acknowledgments}

The first author was supported by an RMIT Vice-Chancellor's PhD
scholarship.
The second and third authors were supported by the Australian
Research Council (grant DP190101113).

\section*{Code}

Code and query variations to reproduce experiments are available 
at \url{https://github.com/rmit-ir/rmitb-trec-covid}.

\bibliographystyle{abbrvnat}
{
 
}

\end{document}